# Magnetic Dynamics of Phase Separation Domains in $GdMn_2O_5$ and $Gd_{0.8}Ce_{0.2}Mn_2O_5$ Multiferroics

V. A. Sanina*, B. Kh. Khannanov, and E. I. Golovenchits

Ioffe Institute, St. Petersburg, 194021 Russia

*e-mail: sanina@mail.ioffe.ru

Specific features of the magnetic properties and magnetic dynamics of isolated phase separation domains in $GdMn_2O_5$ and $Gd_{0.8}Ce_{0.2}Mn_2O_5$ have been investigated. These domains represent 1D superlattices consisting of dielectric and conducting layers with the ferromagnetic orientation of their spins. A set of ferromagnetic resonances of separate superlattice layershas been studied. The properties of the 1D superlattices in $GdMn_2O_5$ and $Gd_{0.8}Ce_{0.2}Mn_2O_5$ are compared with the properties of the previously investigated $RMn_2O_5$(R = Eu, Tb, Er, and Bi) series. The similarity of the properties for all the $RMn_2O_5$ compounds with different R ion types is established. Based on the concepts of the magnetic dynamics of ferromagnetic multi-layers and properties of semiconductor superlattices, a 1D model of the superlattices in $RMn_2O_5$ is built.

1. INTRODUCTION.

$RMn_2O_5$ manganites, where R is the rare-earth ions, Y and Bi, belong to the class of type-II multiferroics, in which the ferroelectric ordering (the Curie temperature is $T_C \approx$ 30–35 K) is induced by the magnetic order (the Neel temperature is $T_N \approx$ 35–45 K) lowering the crystal symmetry to noncentral. Due to the similarity of the $T_C$ and $T_N$ values, the $RMn_2O_5$ manganites exhibit the pronounced magnetoelectric effect, which evokes great interest of researchers in these materials [1, 2]. Until recently, it has been commonly accepted that at $T < T_C$ the $RMn_2O_5$ manganites are characterized by the centrosymmetric sp. gr. Pbam. When these crystals contain the same number of ions of different valences (Mn3+ and Mn4+), it becomes energetically more favorable to establish a certain order (charge order) in their distribution. Mn3+ and Mn4+ ions characterized by alternating ferro- and antiferromagnetic spin orientations at temperatures of $T \leq T_N$ are distributed pairwise along the b axis. The difference between the ferromagnetic double exchange (the exchange integral is $J_{DE} \approx$ 300 meV [3, 4]) and indirect antiferromagnetic exchange ($J_{AF} \approx$ 10 meV) values leads to the exchange striction, which breaks the lattice symmetry, and to the occurrence of ferroelectric ordering along the b axis at $T \leq T_C$ [5].

In the recently published structural work of a series of $RMn_2O_5$ (R = Pr, Nd, Gd, Tb, and Dy) multiferroics, the room-temperature synchrotron resonance scattering study revealed, along with high intensit reflections of the sp. gr. Pbam, the weak-intensity reflections, which could not be described by the central symmetry [6]. Based only on the structural data, the authors of [6] failed to give preference to one of the two possible monoclinic space groups, i.e., P2, which allows polarization along the c axis, and Pm with the polarization in the ab plane. The authors believed that the noncentral groups belong to the entire crystal, in which the uniform polarization can only be established along one crystal axis. Since at low temperatures the striction polarization is oriented along the b axis, the authors chose the sp. gr. Pm. It followed from [6] that at room temperature the $RMn_2O_5$ man-ganites should be characterized by the uniform electric polarization along the b axis, the origin of which is different from that of the low-temperature polarization. The electric polarization in $RMn_2O_5$ (R = Gd and Bi) in the temperature range of 5–330 K along all the crystal axes was studied in [7, 8]. At temperatures above $T_C \approx$ 30–35 K, the electric polarization uniform over the entire crystal was not observed. However, over the entire temperature range, there were local polar domains of phase separation. These regions formed the frozen superparaelectric state, which existed from the lowest temperatures to certain temperatures in the paramagnetic temperature range depending on the crystal axes directions. In these states, the response to the applied electric field had the form of hysteresis loops. This situation was theoretically investigated in [9], but not experimentally observed. The noncentrosymmetric reflections observed in [6] apparently correspond to the polar phase separation domains and the main crystal matrix has the sp. gr. Pbam in the paramagnetic temperature range [7, 8]. The $RMn2O5$ unit cell contains one $Mn^{3+}$ ion and one $Mn^{4+}$ ion. $Mn^{4+}$ ions are located in the nearly undistorted oxygen octahedra (MnO6) and occupy the positions with z= 0.25c and z= (1 – 0.25)c. Their 3d shells contain three localized $t_{2g}$ electrons. $Mn^{3+}$ ions are located in the noncentral pentagonal pyramids (MnO5) in the positions with z= 0.5c and contain three localized $t_{2g}$ electrons and one delocalized $e_g$ electron on the degenerate orbital doublet. $R^{3+}$ ions are located in the positions with z= 0c and have the environment similar to that of $Mn^{3+}$ ions [10]. Thus, the $RMn_2O_5$ manganite is characterized by the layer-by-layer ion distribution perpendicular to the c axis. Transport of $e_g$ electrons between $Mn^{3+}$ and $Mn^{4+}$ ion pairs in the neighboring lattice layers leads to the phase separation, which affects the properties of $RMn_2O_5$ in both the low-temperature multiferroic state and paramagnetic state. This phase separation is analogous to that observed in the doped $LaAMnO_3$ (A=Sr, Ba, and Ca) manganite perovskites, which are magnetic semiconductors with the colossal magneto-resistance [4, 11]. Pure $LaMnO_3$ (sp. gr. Pbnm) is an antiferromagnet with $T_N$= 240 K and Mott insulator. It consists of $Mn^{3+}$ ions in the octahedral oxygen environment. Upon doping of $LaMnO_3$ with $A^{2+}$ ions, manganese ions change their valence ($Ln^{3+} + Mn^{3+} \rightarrow A^{2+} + Mn^{4+}$). As a result, the ion pairs $Mn^{3+}$–$Mn^{4+}$ arise in $LaAMnO_3$ in concentration depending on the degree of doping. The ion pairs $Mn^{3+}$–$Mn^{4+}$ both in $LaAMnO_3$ and $RMn_2O_5$ tend to be accumulated in the ferromagnetic conducting phase separation domains inside the dielectric antiferromagnetic (paramagnetic) crystal matrix due to the strong interactions between them and the crystal matrix [4, 11]. These interactions are the double exchange caused by transport of the $e_g$ electron between the ion pairs $Mn^{3+}$–$Mn^{4+}$ with the ferromagnetic spin orientation of these ions. As was mentioned above, the transfer integral at the double exchange is $t \approx 300$ meV [3, 4]. The double exchange leads to accumulation of $e_g$ electrons in the phase separation domains. This is facilitated also by local structural distortions near the Jahn–Teller $Mn^{3+}$ ions in octahedra ($E_{JT} \approx 700$ meV). The Coulomb repulsion ($E_Q \approx 1$ meV) prevents charge accumulation in these regions. This results in the formation of the dynamical equilibrium localized ferromagnetic conducting phase separation domains. Since the phase separation is induced by the strong interactions, it exists from the lowest temperatures to the temperatures above room temperature. In [4, 11], these regions in $LaAMnO_3$ were considered as nanoscale (~30 nm) spheres. In the later works, it was shown that the phase separation domains arise also in other crystals containing ions of different valences. The shape of these regions depends on the ratio between values of the interactions inducing them and can be 2D plane or 1D [12–16].

The states induced by the phase separation and self-organization of $e_g$ electrons were thoroughly studied in $EuMn_2O_5$ multiferroic manganites and doped $Eu_{0.8}Ce_{0.2}Mn_2O_5$ manganites in [15, 16], where their dielectric and magnetic properties were compared, the specific heat were investigated, and X-ray diffraction and Raman scattering examinations were carried out. The phase separation was observed in the initial crystals, but doping of $EuMn_2O_5$ with $Ce^{4+}$ ions replacing $Eu^{3+}$ ions led to its significant enhancement via forming an additional doping channel. Indeed, in $Eu_{0.8}Ce_{0.2}Mn_2O_5$ in the z= 0 plane, electrons were induced by the reaction $Eu^{3+}= Ce^{4+}+ e$, which transformed $Mn^{4+}$ ions to $Mn^{3+}$ ions in the planes z= 0.25c and 1 – z= 0.75c. The phase separation led to the sharp growth of the permittivity and local conductivity at temperatures above 185 K [15]. Study of the fine structure of Bragg reflections by the high-sensitivity three-crystal X-ray diffraction technique revealed the layered superstructure in $EuMn_2O_5$ and $Eu_{0.8}Ce_{0.2}Mn_2O_5$ along the c axis at room temperature, which consists of the alternating initial $EuMn_2O_5$ dielectric crystal layers and layers with excess $Mn^{3+}$ ions. The layer widths in the 2D superstructure were ~900 Å for $EuMn_2O_5$ and ~700 Å for $Eu_{0.8}Ce_{0.2}Mn_2O_5$ [15]. At low temperatures, the 2D phase separation layers transformed to the limited 1D superlattices consisting of the alternating conducting and dielectric layers with the ferromagnetically oriented spins. Study of $EuMn_2O_5$ and $Eu_{0.8}Ce_{0.2}Mn_2O_5$ specific heats showed that at low temperatures they undergo the same set of phase transitions [15]. It means that the 1D superlattices occupy a small crystal volume both in the initial and doped states. In the superlattice layers, a set of ferromagnetic resonances (FMRs) was observed, the intensity of which sharply dropped at T ≤30 K [17, 18]. As was shown in [16], at temperatures above 40–50 K, carrier hoppings from the 1D superlattices are intensified and, as the temperature increases, the 2D layered structure forms, which acquires the final shape at 185–320 K [15].

The aim of this study was to compare the magnetic properties of GdMn$_2$O$_5$ and Gd$_{0.8}$Ce$_{0.2}$Mn$_2$O$_5$ and magnetic dynamics of the 1D superlattices (a set of FMRs of their separate layers) at temperatures of 5–90 K, which still have been understudied. The especially interesting RMn$_2$O$_5$ multiferroic is GdMn$_2$O$_5$. Gd$^{3+}$ ions (the ground state is $^8S_{7/2}$) have the maximum spin in the series of R ions and are not coupled to the lattice, in contrast to most magnetic R ions. As was shown in [19–21], in GdMn$_2$O$_5$ at temperatures of T≤30 K = T$_C$, the homogeneous antiferromagnetic state with a wave vector of q= (1/2, 0, 0) is established. In this case, the electric polarization appeared higher than that observed usually in RMn$_2$O$_5$ with other R ions by an order of magnitude, which is related to the polarization enhancement by the strong homogeneous Gd–Mn exchange [19]. In this study, we investigate specific features of the magnetic state and low-temperature phase separation in GdMn$_2$O$_5$ and Gd$_{0.8}$Ce$_{0.2}$Mn$_2$O$_5$, compare their properties with those of the previously investigated RMn$_2$O$_5$ manganites, and establish the nature and main peculiarities of the low-temperature phase separation domains in RMn$_2$O$_5$, which were found to be weakly dependent on the R ion type.

2. EXPERIMENTAL

The GdMn$_2$O$_5$ and Gd$_{0.8}$Ce$_{0.2}$Mn$_2$O$_5$ single crystals were grown by spontaneous crystallization and had the form of plates with a thickness of 1–3 mm and an area of 3–5 mm$^2$. The magnetic measurements were performed on a Quantum Design PPMS magnetometer. The microwave magnetic dynamics was studied on a transmission magnetic resonance spectrometer in the frequency range of 28–40 GHz with the low-frequency magnetic modulation [17, 18].

2.1. Magnetic Properties of GdMn$_2$O$_5$ and Gd$_{0.8}$Ce$_{0.2}$Mn$_2$O$_5$

The RMn$_2$O$_5$ compounds with different R ions usually have a complex magnetic structure with the wave vector q= (1/2 + δx, 0, 1/4 + δz) and undergo a great number of magnetic phase transitions at temperatures of T ≤ T$_N$. During these phase transitions, commensurability parameters δx and δz change step-wise [10]. As the temperature near T$_N$ decreases, the incommensurable magnetic phase is observed. With a further decrease in temperature, this incommensurable phase transforms to the commensurable one. Then, at lower temperatures, the incommensurable phase forms again. The ferroelectric ordering in $RMn_2O_5$ is observed, as a rule, in the intermediate commensurable phase. The $GdMn_2O_5$ magnetic structure differs from the structure observed usually in $RMn_2O_5$. As was mentioned above, the commensurable homogeneous collinear antiferromagnetic structure with the wave vec-tor q= (1/2, 0, 0) is observed in the temperature range of 0–30 K ($T_N$= 35 K) [19–21]. Figure 1 shows temperature dependences of the magnetization of $GdMn_2O_5$ in a magnetic field of H= 5 kOe oriented along different crystal axes. It is noteworthy that the magnetization of $GdMn_2O_5$ is much higher than in many $RMn_2O_5$ compounds, which is reasonable to attribute to the contribution of $Gd^{3+}$ ions with the spin S= 7/2. The anomaly caused by the antiferromagnetic transition of Mn ions near $T_N$ is barely noticeable against the background of large contribution of Gd ions to the total magnetization (see the left inset in Fig. 1). For $Gd^{3+}$ ions, the magnetization maximum is observed along the a axis near 13 K, which evidences for the intrinsic ordering of these ions. The right inset in Fig. 1 shows the temperature dependence of the inverse magnetization of $GdMn_2O_5$, which made it possible to determine Curie–Weiss temperatures $T_{CW}$ along different axes. It was found that these temperatures are almost identical to $T_N$. This means that the magnetic state of $GdMn_2O_5$ is not frustrated and the antiferromagnetic transition is sharp. Usually, the $RMn_2O_5$ compounds with other R ions are strongly frustrated and their Curie–Weiss temperature is $T_{CW} \approx (6–7)T_N$ [10]. Comparison of Figs. 1 and 2 shows that the magnetization of the doped crystal is somewhat lower, but still determined by the strongly magnetic Gd ions. Dilution of $Gd^{3+}$ ions with $Ce^{4+}$ ions leads to the disappearance of the ordering in the Gd subsystem near 13K. In this case, the antiferromagnetic transition of Mn ions is almost invisible against the background of the magnetization of Gd ions, as was observed in the initial crystal. The temperature $T_{CW}$, as in $GdMn_2O_5$, almost coincides with $T_N$ of the initial crystal; i.e., the magnetic state of the doped crystal is not frustrated either.

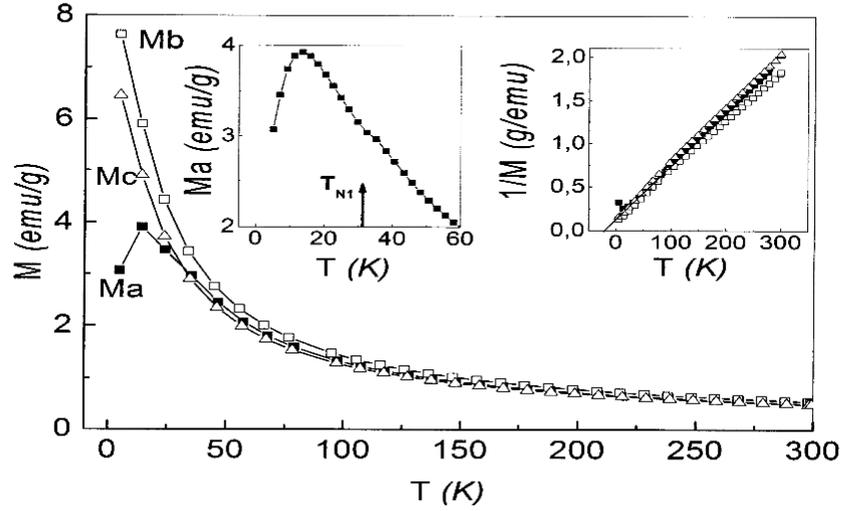

Fig. 1. Temperature dependence of magnetization of $GdMn_2O_5$ along the a, b, and c axes in a magnetic field of 5 kOe. Upper left inset: the same along the a axis in the enlarged scale. Right inset: temperature dependence of inverse magnetization along all the three axes in H=5kOe.

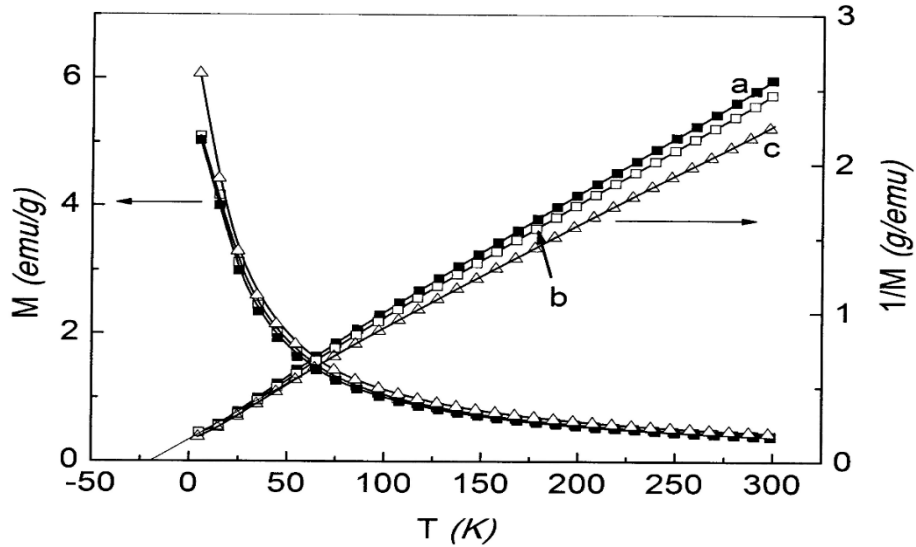

Fig.2. Temperature dependence of magnetization of $Gd_{0.8}Ce_{0.2}Mn_2O_5$ along different crystal axes in a field of H= 5 kOe (left axis) and inverse magnetization (right axis).

## 2.2. Ferromagnetic Resonances of the 1D Layers of the Superlattices: Phase Separation Domains at Low Temperatures

As was mentioned above, the authors of [17, 18] found the 1D superlattices of the alternating conducting and dielectric layers with the ferromagnetic orientation of their spins at low temperatures in a wide series of the $RMn_2O_5$ compounds. Figure 3 shows sets of FMRs detected from separate superlattice layers for these $RMn_2O_5$ compounds. The magnetic dynamics was most thoroughly studied for the phase separation domains in the $EuMn_2O_5$ and $Eu_{0.8}Ce_{0.2}Mn_2O_5$ multiferroics [17]. It appeared that the low-temperature phase separation in $GdMn_2O_5$ and $Gd_{0.8}Ce_{0.2}Mn_2O_5$ is also a set of isolated 1D superlattices located in the initial multiferroic matrix, which occupies a small matrix volume.

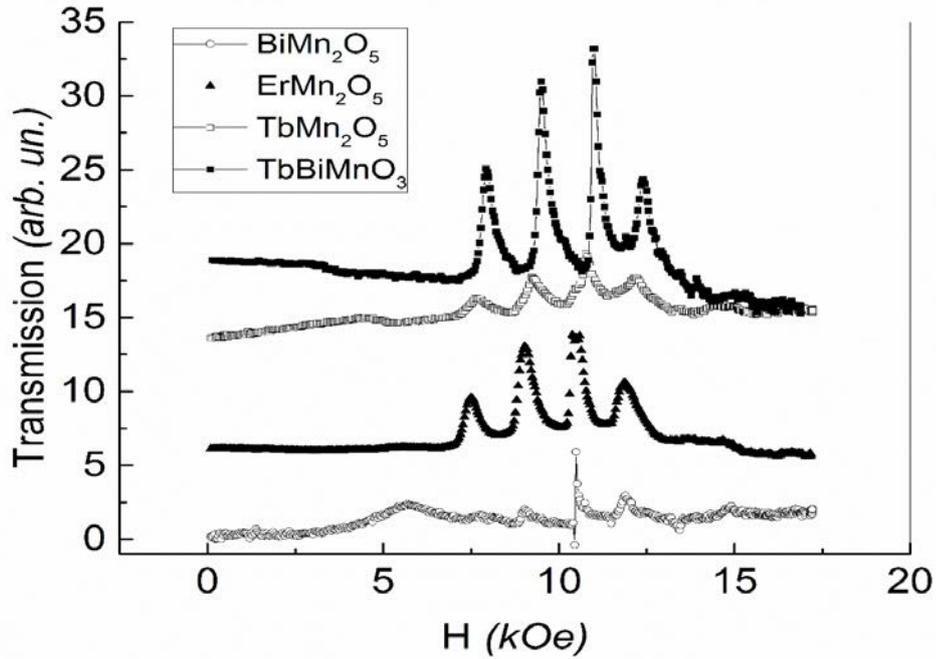

Fig. 3. Sets of ferromagnetic resonances from the 1D superlattice layers for a series of multiferroic manganites. The temperature is T= 5 K and the frequency is F=31.6 GHz.

Figures 4 and 5 show FMR sets for $GdMn_2O_5$ and $Gd_{0.8}Ce_{0.2}Mn_2O_5$, respectively. It can be seen that in both crystals there is the most intense and narrow L0 line in the external magnetic field exactly corresponding to the FMR frequency at the given frequency for an isotropic ferromagnet at g-factor g= 2. The FMR lines with the lower intensity on the left from the fundamental resonance (L1 and L2) and on the right from it (R1 and R2) are positioned almost symmetrically relative to the L0 line in weaker and stronger magnetic fields. The ferromagnetic 1D layers in $GdMn_2O_5$ and $Gd_{0.8}Ce_{0.2}Mn_2O_5$ are isotropic and oriented by the external magnetic field; i.e., they are not coupled to the main crystal matrix. Temperature dependences of separate FMR lines are presented in Figs. 4 and 6. It can be seen that the set of FMR lines is observed only at low temperatures (below 30 K in $GdMn_2O_5$ and below 40 K in $Gd_{0.8}Ce_{0.2}Mn_2O_5$ for all the lines, except for R2). The FMR sets for GdMn$_2$O$_5$ and Gd$_{0.8}$Ce$_{0.2}$Mn$_2$O$_5$ are similar, but the intensities of lines in Gd$_{0.8}$Ce$_{0.2}$Mn$_2$O$_5$ are much higher. This evidences for the higher concentration of the phase separation domains in Gd$_{0.8}$Ce$_{0.2}$Mn$_2$O$_5$.

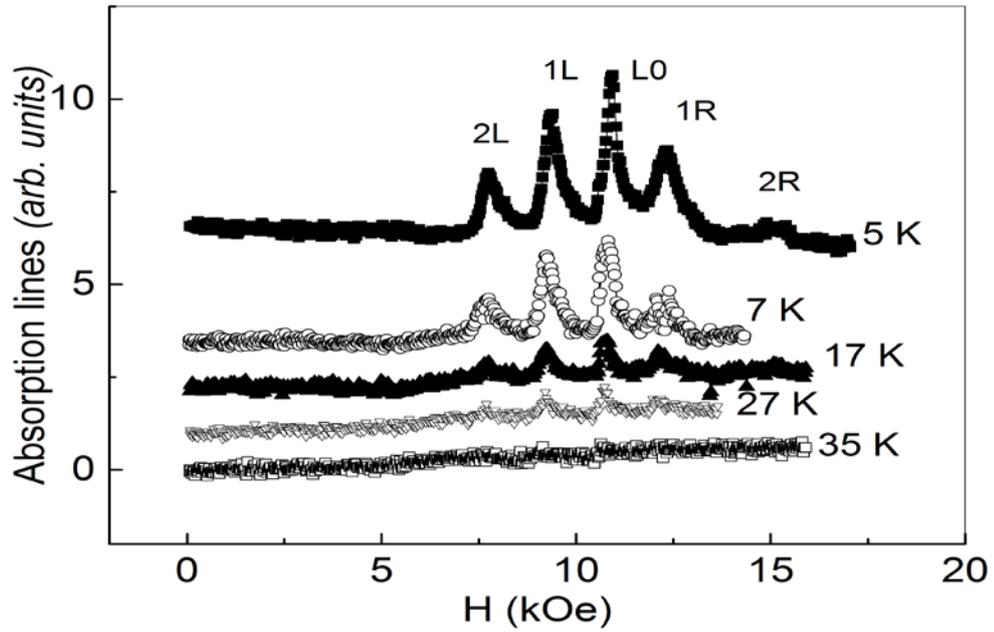

Fig. 4. Set of FMR lines from separate 1D superlattice layers for GdMn$_2$O$_5$ upon magnetic field sweep H||c at some temperatures.

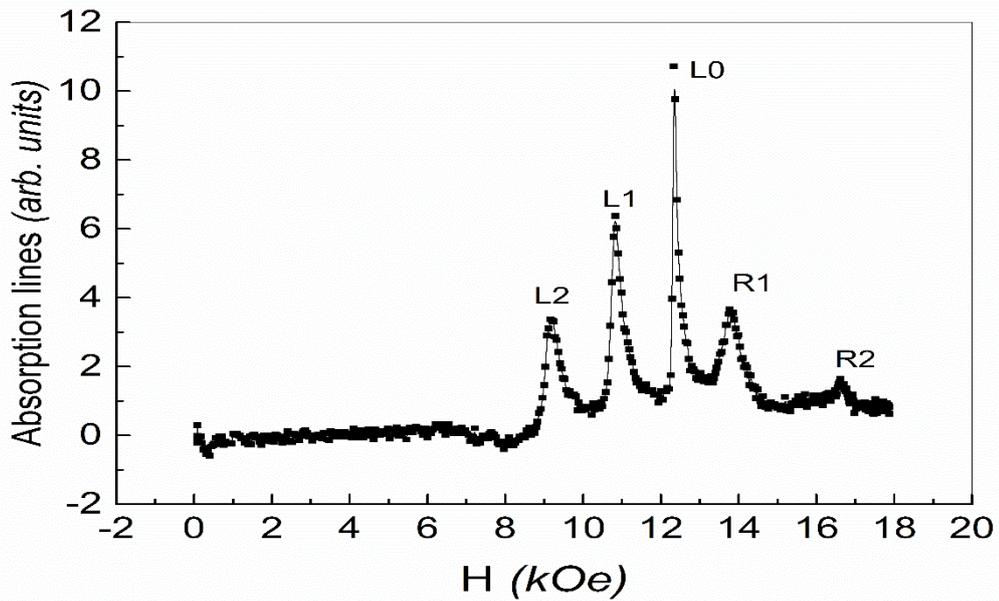

Fig. 5. Set of FMR lines from separate 1D superlattice layers for Gd$_{0.8}$Ce$_{0.2}$Mn$_2$O$_5$ upon magnetic field sweep H||c at a frequency of 33.16 GHz at T= 5 K.

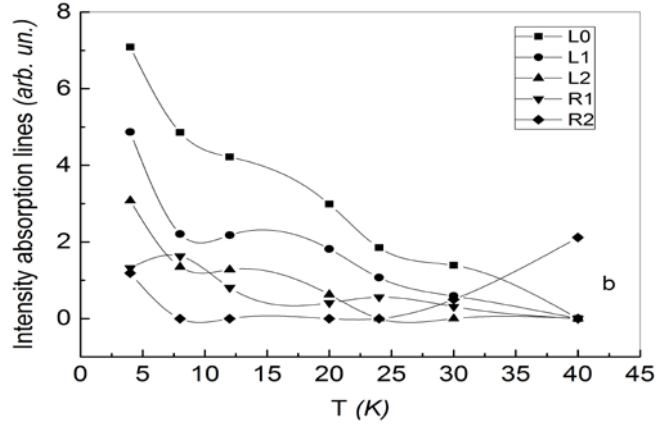

Fig. 6. Temperature dependences of intensities of a set of FMR lines for $Gd_{0.8}Ce_{0.2}Mn_2O_5$.

Figures 7 and 8 present frequency dependences of the resonance magnetic fields for a set of $GdMn_2O_5$ and $Gd_{0.8}Ce_{0.2}Mn_2O_5$ lines, respectively. One can see the linear magnetic-field dependences of the FMR line frequencies. These dependences can be described as $\varpi_n = \gamma_n (H+H_{eff}^n)$. Here, $\gamma_n$ is the gyro-magnetic ratio for the n lines (n= L1, L2, L0, and R1), $H_{eff}^n$ is the internal effective field forming a gap in the $\varpi_n(H)$ dependence, and H is the external magnetic field. The values $H_{eff}^n$ are positive for the L1 and L2 lines and increase for these lines with line number L. The values $H_{eff}^n$ for the R1 lines in both crystals are negative. The g-factors for the R1 lines are minimum and somewhat smaller than 2. The g-factors for the Ln lines are close to 2.

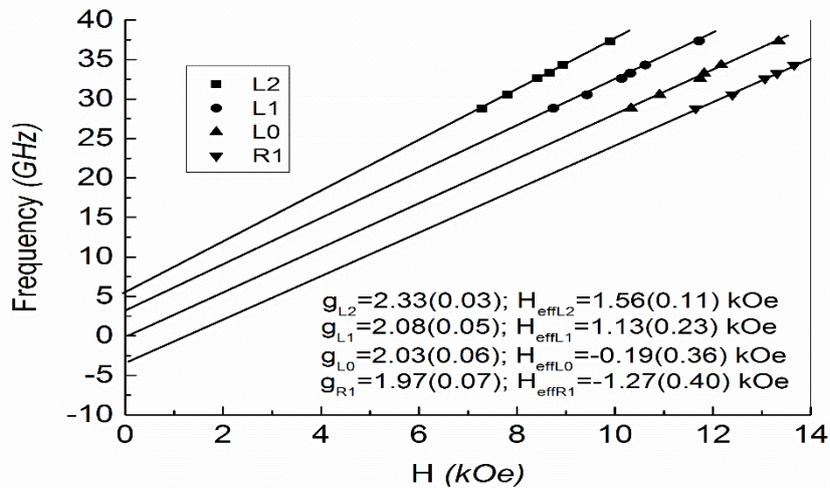

Fig. 7. Frequency dependences of resonant magnetic fields for all the FMR lines in $GdMn_2O_5$ at 5 K. In the field of the picture, the g-factors and values $H_{eff}^n$ for separate lines are presented.

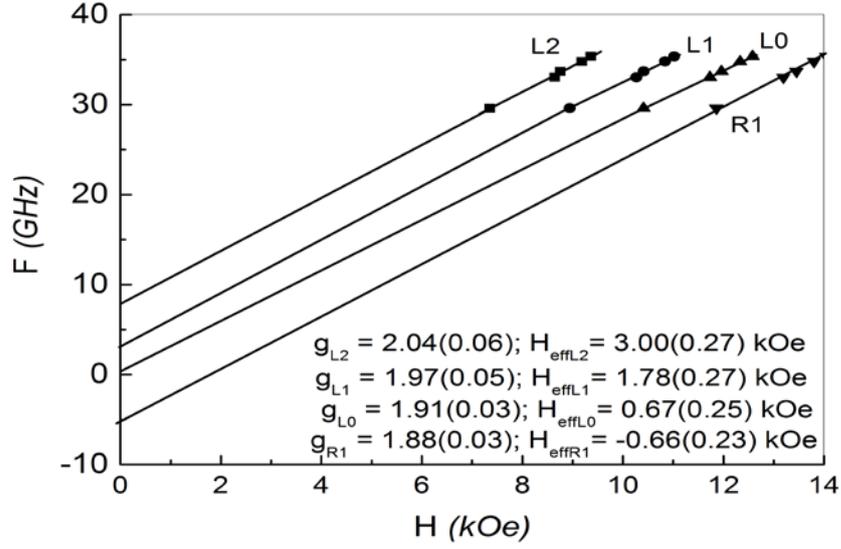

Fig. 8. Frequency dependences of resonant magnetic fields for the FMR lines in $Gd_{0.8}Ce_{0.2}Mn_2O_5$ at 5 K. In the field of the picture, the g-factors and values $H_{eff}^n$ for separate lines are presented.

Comparison of the magnetic fields in which separate FMR lines are observed and the g-factors for these lines shows that they differ insignificantly in the $RMn_2O_5$ compounds with different R ions and in the initial and doped crystals (Figs. 3–5). In this case, Er3+ and Tb3+ ions are strongly coupled to the lattice, while Gd3+ ions are almost uncoupled with it. It means that the values $H_{eff}^n$ are not specified by the anisotropy fields, which should be determined by the R ion type inside the layers, but are caused by the internal fields exceeding by far the anisotropy fields. Indeed, as was mentioned above, the dynamical equilibrium state of the phase separation domains is determined by a balance of the strong interactions. The differences between the parameters of separate superlattice lines are most likely caused by the difference in the distributions of $Mn^{3+}$ and $Mn^{4+}$ ions in the superlattice layers occurring upon $e_g$ electron transport between these pairs of ions. As a result, a set of layers of the 1D superlattices can be presented in the form of periodically changing isotropic ferromagnetic layers.

To describe the spin-wave excitations in the superlattice layers, we can use the dispersion equation for spin waves of isotropic magnetic films in multilayers [22]

$$\omega^2 =(\omega_H +\eta k^2) (\omega_H +\eta k^2 +\omega_M \sin \theta_k^2) \qquad (1)$$

Here $\omega_H = 4\pi\gamma H$, $\omega_M = 4\pi\gamma M_0$, $\eta$ is the inhomogeneous exchange constant, $\theta_k$ is the angle between directions of wave vector k of the spin wave and magnetic field H, and $M_0$ is the saturation magnetization. In the case of longitudinal magnetization ($\theta_k = 0$) or at $\omega_H \gg \omega_M$, at the transverse magnetization implemented in our case, the dispersion law for spin waves is $\omega = \omega_H + \eta k^2$. In addition, we should take into account the boundary conditions caused by spin pinning at the boundaries of separate superlattice layers. The pinning parameter $\xi$ is a ratio between the energy of surface anisotropy $K_S$ of the layer and the inhomogeneous exchange energy $(qM_0)^2$ in it, i.e., $\xi = 2K_S/(qM_0)^2$. The pinning parameters $\xi_1$ and $\xi_2$ on the opposite surfaces determine wave vector k of the standing spin wave excited in a layer with thickness d. This relation is described by the equation [22]

$$\coth(kd) = (k^2 - \xi_1\xi_2)/[k(\xi_1+\xi_2)]. \qquad (2)$$

In our case, the dynamical equilibrium states of the superlattice are established and maintained by permanent hoppings of $e_g$ electrons between $Mn^{3+}$ and $Mn^{4+}$ ions both inside the layers and between separate layers. As was mentioned above, such hoppings can occur only when Mn ion spins are oriented ferromagnetically in all the neighboring layers, which is caused by the double exchange. As a result, the condition $\xi_1 = \xi_2 = 0$ should be met for all the inner layers of the superlattices. In this case, as follows from Eq. (2), the spin-wave excitations with the wave vectors $k_m = (m-1)\pi/d$ (m = 1, 3, 5…) can occur in the inner layers of the superlattices. Then, the fundamental modes with m=1 are the homogeneous spin-wave excitations with k=0, i.e., homogeneous ferromagnetic resonances. The higher-order modes with m = 3, 5, …, etc., are the standing spin waves with k ≠ 0. At a fixed frequency, such modes, which exhibit decreasing intensity with increasing m, should be excited in the decreasing magnetic fields due to the increasing contribution of the inhomogeneous exchange $\eta k^2$. In the 1D superlattices, we apparently only observe the fundamental modes of the homogeneous FMR in all superlattice layers. The rigid spin pinning can occur in the investigated superlattices only on the outer interfaces between the edge layers and the antiferromagnetic crystal matrix. Near such interfaces, the inhomogeneous internal magnetic field is induced, which changes the wave vector k of the spin wave. The outer interface is a rotation surface (Fig. 9), on which wave vector k of the spin wave changes its sign and passes through zero. It is positive inside the superlattice and negative outside it. This means that the homogeneous spin excitation with alternating magnetic moment m propagates inside the superlattice layer, but decays exponentially in the antiferromagnetic matrix. As a result, the spin excitation with k= 0 (Fig. 9) occurs also on the rotation surface [22]. Thus, the homogeneous spin-wave excitations with k= 0 (FMR) are induced in all the superlattice layers and on their rotation surfaces. The presence of the rotation surfaces near the superlattice edges makes the ferromagnetic superlattices independent of the main antiferromagnetic matrix and their spin moments are oriented by only external field H.

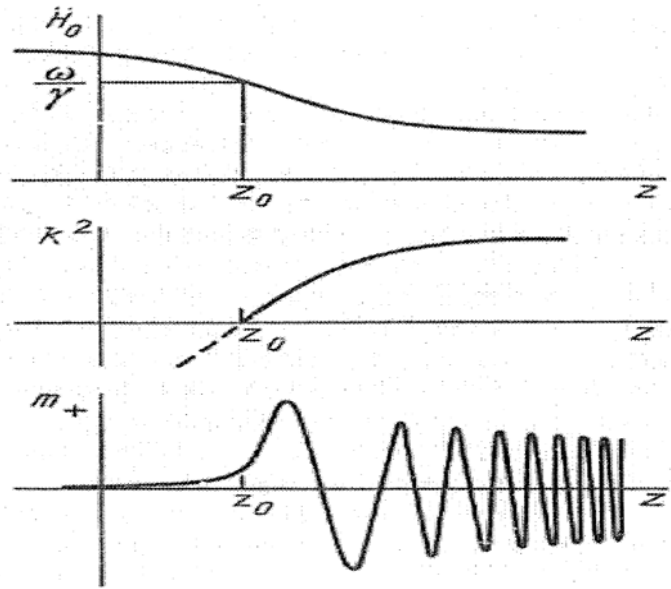

Fig. 9. Inhomogeneity of field $H_0$ and wave vector $k^2$ of the spin wave and propagation of the spin wave with variable magnetization m near the rotation surface (at the boundary surface of the superlattice layer with coordinate z=0) [22].

The aforementioned features of the spin excitations in the investigated superlattices are caused by the fact that they are formed in the bulk of multiferroic single crystals by self-organization. The condition $\xi_1 = \xi_2 = 0$ cannot be met in artificial ferromagnetic superlattices due to the inevitable defect formation at the interfaces between different layers. In such multilayers, the higher-order inhomogeneous modes with k≠0 are usually excited. As was mentioned above, such modes with decreasing intensity are localized in decreasing magnetic fields from the most intense mode. Only the surface modes in artificial ferromagnetic multilayers are excited in strong magnetic fields with respect to the fundamental line, but usually close to it. In this case, the surface mode intensities can exceed the intensity of the fundamental mode in the film layer [22]. For artificial superlattices, the probability of observation of the almost symmetric higher-order modes and surface modes with the similar intensities is almost zero. Thus, we only observe the fundamental modes with k= 0 from different superlattice layers with the maximum intensity, which are nearly symmetric relative to the most intense L0 line.

To explain this symmetry, assume the 1D superlattice layers contain $Mn^{3+}$ and $Mn^{4+}$ ions in different ratios and form a semiconductor heterostructure. Then, $Mn^{3+}$ and $Mn^{4+}$ ions in these layers are donors and acceptors, respectively. The L0 layer contains the same number of these ions of different valences, the charge ordering of which make these layers dielectric. The fully compensated semiconductor L0 layer forms. In this case, the Fermi level in this layer is located at the band gap center [23]. This layer has the maximum intensity and minimum FMR linewidth among other lines. In the L0 layers, FMR is observed at a frequency of $\omega_0 = \gamma_0 H$, $H_{eff}^n \approx 0$. In the R1 layers, the values $H_{eff}^n$ are negative, which can be explained by the fact that the number of $Mn^{3+}$ ions in this layer exceeds the number of $Mn^{4+}$ ions and there are excess $e_g$ electrons. The excess Mn3+ ions are located in the positions of Mn4+ ions (in the oxygen octahedra), i.e., represent Jahn–Teller ions and lead to the local lattice distortions. This results in the distortion of the strain potential inside the layer; the deepest wells intersect the Fermi level and form electron droplets [23]. The diamagnetic contribution of magnetization of electrons of these droplets in applied magnetic field H forms the negative gap. In addition, excess electrons reduce the g-factors of the R1 lines to a value of smaller than 2 and broaden the R1 lines. On the contrary, in the L1 and L2 layers, we observe the dominance of $Mn^{4+}$ ions and hole conductivity caused by the much lower concentration of $Mn^{3+}$ ions and $e_g$ electrons. The positive values $H_{eff}^n$ are caused by paramagnetism of localized spins of excess $Mn^{4+}$ ions. In addition, the FMR lines are broadened by scattering of spin excitations on hole carriers.

The charge difference L–0–R between the neighboring layers leads to the local electric polarization of the 1D superlattices. In [24], we built a schematic model of 1D superlattices in the form of the alternating L–0–R layers, which explains the occurrence of local electric polarization in $Eu_2CuO_4$ consistent with the situation observed in $RMn_2O_5$ (Fig. 10). The difference between the numbers of $Mn^{3+}$ and $Mn^{4+}$ ions and carriers in the L–0–R layers leads to the occurrence of the resulting polarization of 1D superlattices—the low-temperature phase separation domains. This polarization was observed by us in $GdMn_2O_5$ at low temperatures along all the crystal axes [7, 8]. Thus, the dynamical equilibrium 1D superlattices occurring at low temperatures in a wide series of $RMn_2O_5$ multiferroic manganites containing manganese ions of different valences have the intrinsic electric polarization of different nature than the striction polarization of the main matrix. As a matter of fact, the L and R layers are a particular case of charged domain walls forming in ferroelectrics between domains with the tale-to-tale or head-to-head electric polarization orientations, which are formed by the L0 layers (Fig. 10) [25, 26].

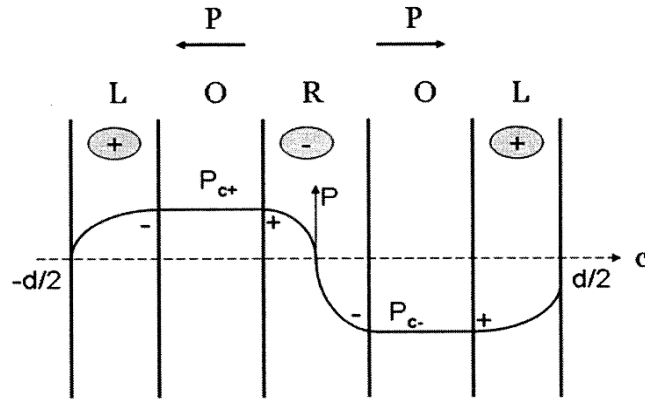

Fig. 10. Schematic of the 1D structure of the superlattice with thickness d and distribution of polarization P in the alternating layers. Closed ellipses show coupled charges (dominance of Mn3+ or Mn4+ ions in the R and L layers, respectively). Screening free carriers are shown by pluses and minuses for holes and electrons, respectively.

The symmetry of the L and R lines relative to the L0 line is caused by the requirement for electroneutrality of the superlattice in the presence of different charges of Mn ions and electron and hole carriers in the right and left superlattice layers (Fig. 10).

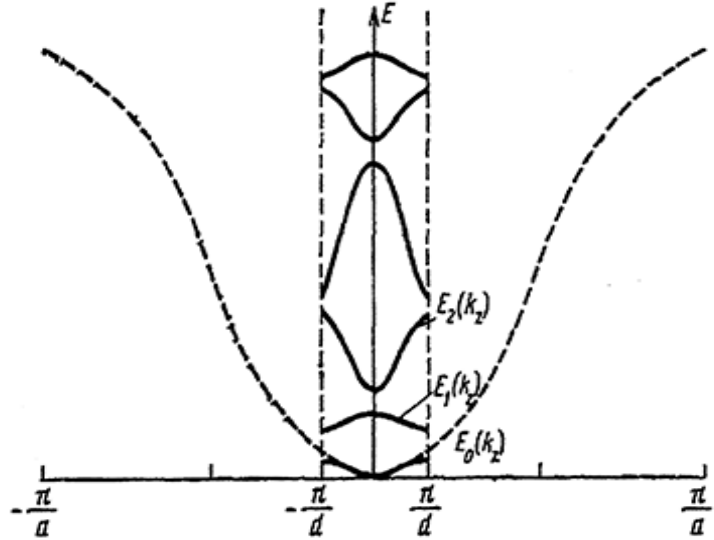

Fig. 11. Splitting of the energy band $E(k_z)$ of the crystal with lattice constant a into the minibands $E_f(k_z)$ by the potential of a superlattice with period d. The number of minizones is d/a [27].

In terms of the semiconductor band diagrams, the superlattices in multiferroics are similar to ordinary semiconductor superlattices and represent a set of split minibands (Fig. 11) [27] with gaps between them. Figure 12 shows the frequency dependence of the observed FMR intensity (the L0 lines) in GdMn2O5 at T= 5 K, which reveals minibands of the 1D superlattices and gaps between them.

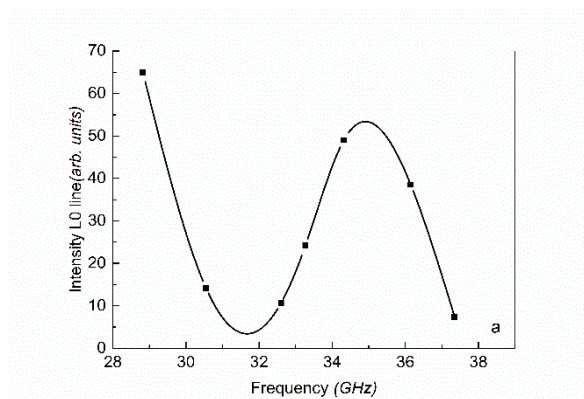

Fig.12. The frequency dependence of FMR intensity (the L0 lines) in GdMn2O5 at T= 5 K.

As the temperature increases, upon breaking of the antiferromagnetic ordering and simultaneous growth of the hopping conductivity, the 1D superlattices change their shape and state. With an increase in the kinetic energy of electrons, the layers with deeper barriers (the R layers) are populated in a greater degree. The number of superlattice layers decreases and the shape of phase separation domains changes, transforming to the 2D structure observed by us at room temperature in $GdMn_2O_5$ [7, 8] and $Gd_{0.8}Ce_{0.2}Mn_2O_5$ [28].

3. CONCLUSIONS

Thus, the dynamical equilibrium 1D superlattices occur at low temperatures in a wide series of $RMn_2O_5$ multiferroic manganites containing manganese ions of different valences. The similarity of the properties of superlattices in $RMn_2O_5$ with R = Gd, Er, Tb, and Eu is indicative of the fact that these properties are almost independent of the R ion type and R–Mn exchange. The properties of the superlattices are caused by the correlations between the spin and charge ordering of manganese ions under strong interactions induced by transport of $e_g$ electrons between $Mn3+$ and $Mn4+$ ions. These superlattices are phase separation domains below the temperatures of existence of the multiferroic state. These superlattices represent localized regions of the alternating conducting and dielectric layers with the ferromagnetic orientation of their spins. Due to the presence of the rotation surfaces at the boundaries of 1D superlattices, the latter are uncoupled with the main antiferromagnetic crystal matrix and their spins are oriented by the external magnetic field.